\documentclass[osajnl,twocolumn,showpacs,superscriptaddress,10pt,floatfix]{revtex4-1} 
\usepackage{amsmath,amssymb,graphicx}
\usepackage{units,siunitx}
\usepackage{lipsum}
\usepackage{braket}
\begin{document}

\title{Experimental generation of an optical field with arbitrary spatial coherence properties} 

\author{Brandon~Rodenburg}
\email{Corresponding author: Brandon.Rodenburg@gmail.com}
\author{Mohammad~Mirhosseini}
\author{Omar~S.~Maga\~{n}a-Loaiza}
\affiliation{The Institute of Optics, University of Rochester, Rochester, New York 14627, USA}

\author{Robert~W.~Boyd}
\affiliation{The Institute of Optics, University of Rochester, Rochester, New York 14627, USA}
\affiliation{Department of Physics, University of Ottawa, Ottawa ON K1N 6N5, Canada}

\begin{abstract}
    We describe an experimental technique to generate a quasi-monochromatic
    field with any arbitrary spatial coherence properties that can be described
    by the cross-spectral density function, $W(\mathbf{r_1},\mathbf{r_2})$.
    This is done by using a dynamic binary amplitude grating generated by a
    digital micromirror device (DMD) to rapidly alternate between a set of
    coherent fields, creating an incoherent mix of modes that represent the
    coherent mode decomposition of the desired $W(\mathbf{r_1},\mathbf{r_2})$.
    This method was then demonstrated experimentally by interfering two plane
    waves and then spatially varying the coherence between them. It is then
    shown that this creates an interference pattern between the two beams whose
    fringe visibility varies spatially in an arbitrary and prescribed way.
\end{abstract}

\ocis{030.0030, 030.1640, 030.4070, 090.1760, 050.1970, 070.6120.}

\maketitle

\section{Introduction}
The transverse degree of freedom of an optical field is the fundamental aspect
of light that contains spatial information. Utilization of this information is
the basic resource in traditional imaging systems and in applications such as
microscopy, lithography, holography or metrology. In addition, use of the
transverse modes of light has recently been demonstrated to be an important
resource in optical
communication~\cite{Wang2012,Rodenburg2012,Mirhosseini2013,Boyd2011},
high-dimensional entanglement studies~\cite{Mair2001a,Dada2011}, and Quantum
key distribution~\cite{Boyd2011a,Malik2012}.

Having control of the spatial coherence properties of a light beam provides an
additional degree of control compared to using fully coherent light only, and
has been shown to be advantageous for a number of applications. Beams of
decreased coherence allow access to spatial frequencies that are twice those
available in a purely coherent system~\cite{Considine1966}. Greater spatial
frequencies can enable improvements in imaging based systems and has been shown
to be particularly useful in lithography~\cite{Lin1980}. Partial coherence also
allows for the suppression of unwanted coherent effects by decreasing the
coherence, such as suppression of speckle~\cite{Dainty1970} which enables lower
noise and opens the door to novel imaging modalities~\cite{Dubois1999}. It has
also been suggested that partial coherence can improve the deleterious effects
of optical propagation through random or turbulent media~\cite{Gbur2002}. In
addition, the coherent property of optical beams can be used for novel beam
shaping~\cite{Lajunen2011} as well as a method for control over soliton
formation due to modulation instabilities in the study of nonlinear beam
dynamics~\cite{Chen2002}. The ability to generate arbitrary optical beams could
also be used as a tool in basic research, such as in optical
propagation~\cite{Waller2012} or testing of novel methods in quantum state
tomography dealing with the transverse wavefunction of light that has seen a
great deal of interest recently~\cite{Lundeen2011,Lundeen2012}. Traditional
methods used to generate partially coherent beams of light often rely on
imprinting a changing pattern of random phase or speckle onto a coherent beam,
such as with a spatial light modulator (SLM)~\cite{Rickenstorff2012} or
rotating diffuser~\cite{Baleine2004}. It has even been demonstrated that SLMs
allow the statistics of the speckle patterns to be varied across the beam to
give spatially varying coherence properties~\cite{Waller2012}. However none of
these methods have been shown to allow for complete arbitrary control over the
spatial coherence of an optical beam.

In this paper we demonstrate how to generate any arbitrary quasi-monochromatic
partially coherent field that can be specified by a cross-spectral density
function $W(\mathbf{r_1,r_2})$, i.e. for fields fully specified by their two
point spatial correlations. This is done by first computing the coherent mode
decomposition of $W(\mathbf{r_1,r_2})$, which is an incoherent mixture of
orthogonal coherent modes. For each of these coherent modes a computer
generated hologram (CGH) is computed for a digital micromirror device (DMD)
that acts as a binary amplitude spatial light modulator with rapid modulation
speeds. The DMD then switches between each coherent mode on timescales slower
than the coherence time of the source laser, but long relative to the detection
time of the CCD. This creates an incoherent averaging that physically
reproduces the coherent mode decomposition.
Section~\ref{sec:CoherentModeDecomposition} details computation of the coherent
mode decomposition, section~\ref{sec:BinaryGratings} describes the algorithm to
compute binary amplitude CGHs for the generation of coherent modes and
section~\ref{sec:Experiment} details the experimental demonstration of this
technique.

\section{Coherent mode decomposition}\label{sec:CoherentModeDecomposition}
The transverse wavefront of a deterministic and coherent scalar beam is
described by a complex field, $U(\mathbf{r})$. For a stochastic beam,
$U(\mathbf{r})$ is a random variable and it becomes necessary to represent the
field in a more sophisticated way. The standard way of doing this is with the
cross-spectral density function. At a single frequency the cross-spectral
density function is defined as
\begin{equation}
    W(\mathbf{r_1},\mathbf{r_2}) = 
    \left<U^*(\mathbf{r_1})U(\mathbf{r_2})\right>,
    \label{eq:coherencefunction}
\end{equation}
and represents the average intensity ($\Braket{I(r)}\equiv W(\mathbf{r,r})$),
as well as the correlations (up to second order) of such a partially coherent
field~\cite{Mandel1995}.

$W(\mathbf{r_1},\mathbf{r_2})$ can be decomposed into an incoherent sum of
orthogonal spatial modes $\psi_n(\mathbf{r})$, written as
\begin{equation}
    W(\mathbf{r_1},\mathbf{r_2}) =
    \sum_n{\lambda_n\psi_n^*(\mathbf{r_1})\psi_n(\mathbf{r_2})},
    \label{eq:decomposition}
\end{equation}
where $\lambda_n$ are real and nonnegative, and $p_n = \lambda_n/\sum\lambda_n$
is the relative weight of the field in mode
$\psi_n(\mathbf{r})$~\cite{Wolf1981}. The modes $\psi_n(\mathbf{r})$ can be
computed as the eigenfunctions with corresponding eigenvalues $\lambda_n$ from
the Fredholm integral equation
\begin{equation}
    \int{W(\mathbf{r_1},\mathbf{r_2})\psi_n(\mathbf{r_1})\,\mathrm{d}^2\mathbf{r_1}}
    =\lambda_n\psi_n(\mathbf{r_2}).
\end{equation}

This representation is often referred to as a coherent mode decomposition of
$W(\mathbf{r_1},\mathbf{r_2})$. Mathematically Eq.~\ref{eq:decomposition} is a
sum over an infinite number of modes, but in practice $n$ is bounded by the
maximum spatial frequency content of $W(\mathbf{r_1,r_2})$, i.e. there is some
maximum $n_{max} = N$ such that for $n > N$, $p_n$ will be negligibly small.
For example, Gaussian Schell-model beams are a common example of a partially
coherent beam. Such a beam is defined by having a Gaussian intensity
$I(\mathbf{r}) = \exp{(-r^2/2\sigma_I^2)}$, as well as a Gaussian degree of
coherence $\mu(\mathbf{r_1,r_2}) =
\exp{(-|\mathbf{r_1-r_2}|^2/2\sigma_\mu^2)}$, which gives a cross-spectral
density function
\begin{equation}
    W(\mathbf{r_1,r_2}) = \sqrt{I(\mathbf{r_1})I(\mathbf{r_2})}\mu(\mathbf{r_1,r_2}).
    \label{eqn:GaussianSchell}
\end{equation}
A coherent mode decomposition of such a Gaussian Schell-model beam shows that
the number of coherent modes necessary to describe Eq.~\ref{eqn:GaussianSchell}
is given by the number of independent coherent regions within the beam which is
quantified by $N\approx (\sigma_\mu/\sigma_I)^2$~\cite{Starikov1982}.

Physically Eq.~\ref{eq:decomposition} can be realized if one can create a beam
that alternates between the coherent modes $\psi_n(\mathbf{r})$ in time with
relative frequency weighted by $p_n$. For measurement to yield the intended
field, the switching time $\tau_s$ must be much faster than any detector
integration time $\tau_{det}$ in order to create the intended averaging over
the inputs. In addition, for the mixture to be an incoherent mixture, the
various modes must not have any correlations in time. Thus the switching time
must be slower than the coherence time $\tau_{coh}$ of the source. Together
these form the condition
\begin{equation}
    \tau_{det} > \tau_s > \tau_{coh}.
    \label{eqn:CoherenceTimes}
\end{equation}
If Eq.~\ref{eqn:CoherenceTimes} is met, then one has a physically realized
implementation of $W(\mathbf{r_1,r_2})$.

\section{Generating Arbitrary Coherent Fields with Binary
Gratings}\label{sec:BinaryGratings}

In order to generate an arbitrary partially coherent field
$W(\mathbf{r_1,r_2})$, one only needs to find a way to create the coherent
fields $\psi_n(\mathbf{r})$ in rapid succession. Such rapid mode generation was
recently demonstrated by using DMDs to create quickly addressable binary
amplitude modulated CGHs~\cite{Mirhosseini2013a}, though this comes at the cost
of having a maximum efficiency around 10\%. DMDs are devices that provide
both the speed and resolution desired for rapid generation and switching of
coherent fields~\cite{Dudley2003}. A DMD consists of a 2-dimensional array of
mirrors that can be in one of two positions, which can be used to act as an on
or off state at each pixel. Each pixel can be individually addressed and
changed very rapidly, at frame rates exceeding $10\,\mathrm{kHz}$.

\begin{figure}[htpb]
    \centering
    \def\svgwidth{0.47\textwidth}
    \setlength{\unitlength}{\svgwidth}
    \begin{picture}(1,0.35)
        \put(0,0){\includegraphics[width=\unitlength]{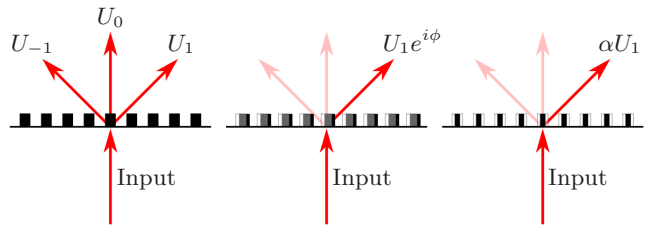}}
        \put(0.16937818,0.07624671){\makebox(0,0)[lb]{\smash{Input}}}
        \put(0.15969989,0.33046017){\makebox(0,0)[b]{\smash{$U_0$}}}
        \put(0.24935869,0.28885149){\makebox(0,0)[lb]{\smash{$U_1$}}}
        \put(0.07004105,0.28885149){\makebox(0,0)[rb]{\smash{$U_{-1}$}}}
        \put(0.50968643,0.07624671){\makebox(0,0)[lb]{\smash{Input}}}
        \put(0.58924682,0.28885149){\makebox(0,0)[lb]{\smash{$U_1e^{i\phi}$}}}
        \put(0.8499949,0.07624705){\makebox(0,0)[lb]{\smash{Input}}}
        \put(0.92955516,0.28885149){\makebox(0,0)[lb]{\smash{$\alpha U_1$}}}
    \end{picture}
    \caption{Left: A binary amplitude grating composed of a series of
        rectangular pulses diffracting light into multiple orders. Middle:
        Pulse position modulation where a phase change is induced in the
        diffracted order as a result of a shift in the pulses.  Right:
        Change in the amplitude of the diffracted order by pulse width
        modulation in which the diffraction efficiency is varied by changing
        the duty cycle of the binary pulses.}
    \label{fig:BinaryGrating}
\end{figure}

The fact that DMDs have 2 settings, allows us to make a binary grating. Any
periodic structure acts as a diffraction grating. A transverse shift in this
diffraction grating will induce a phase shift or detour phase in the diffracted
orders, even if the grating is an amplitude only structure. In addition, the
form of each period will determine the scattering efficiency into the
diffracted order.  Taken together, modulating the grating position and each
periodic form locally within the hologram allows one to control both the
amplitude and phase, and thus create any field, $U(\mathbf{r}) = A(\mathbf{r})
\exp{\left(i\phi(\mathbf{r})\right)}$ in the diffracted order.

A well known method of encoding binary holograms is to create a periodic array
of binary fringes or rectangular `pulses.' A one dimensional representation of
this is shown in Fig.~\ref{fig:BinaryGrating}. A shift in the location of these
pulses will change the overall phases into the diffracted orders, while
changing the widths or duty cycles of the pulses will change the diffracted
efficiency. These two methods are known as pulse position and pulse width
modulation respectively~\cite{Brown1969,Lee1979,Mirhosseini2013a}, and such a
modulation represents a generalization of the Moir\'e
technique~\cite{Zhang2011}. Mathematically, a periodic binary grating can be
written as a Fourier series
\begin{equation}
    f(\mathbf r) = \sum_m \frac{\sin{(\pi mq)}}{\pi m}e^{im(\mathbf{G\cdot r} + 2\pi\delta )},
    \label{eqn:GratingFT}
\end{equation}
where $\mathbf G=\frac{2\pi}{T}(\cos(\theta)\mathbf{\hat{x}} +
\sin(\theta)\mathbf{\hat{y}})$ is the grating wavevector. The grating consists of
rectangular pulses of width $qT$ spaced at a period of $T$ and $\delta\in
[-1/2,1/2]$ is the relative location of the array within each period.  Looking
only at the first diffraction order $m=1$ the field is given by
\begin{equation}
    U_1 = U_{in}*\frac{\sin{(\pi q)}}{\pi}e^{i2\pi\delta},
    \label{eqn:DiffractedOrder}
\end{equation}
where $U_{in}$ is the input field, which we'll assume to be a constant plane
wave. In addition all optics after the DMD are aligned along the axis of the
first diffraction order, allowing us to ignore any phase tilt from $U_{in}$ as
well as the $e^{i\mathbf{G\cdot r}}$ tilt from the grating in our description
of $U_1$.

We can allow $q$ and $\delta$ to become functions of position and the previous
results still hold so long as $q(\mathbf{r})$ and $\delta(\mathbf{r})$ vary
much slower than the grating period $T$. Then any complex field
$A(\mathbf{r})e^{i\phi(\mathbf{r})}$ can be generated by allowing
\begin{equation}
    q(\mathbf{r}) = \frac{1}{\pi}\arcsin{\left(A(\mathbf{r})\right)}, \qquad
    \delta(\mathbf{r}) = \frac{\phi{(\mathbf{r})}}{2\pi},
    \label{eqn:PulsePostionAndWidth}
\end{equation}
where the phase is $\phi\in[-\pi,\pi],$ which is defined symmetrically around 0
to avoid encoding errors in the presence of a varying
amplitude~\cite{Bolduc2013}.

This full procedure can be represented in the following fashion. First one
chooses the field $U=A(\mathbf{r})e^{i\phi(\mathbf{r})}$ that one wishes to
create. Then $q(\mathbf{r})$ and $\delta(\mathbf{r})$ are computed from
Eq.~\ref{eqn:PulsePostionAndWidth} and a periodic sinusoidal function is
computed to give
\begin{equation}
    \cos\left(\mathbf{G\cdot r} + 2\pi \delta(\mathbf{r})\right).
\end{equation}
To convert this into a binary hologram, this function is thresholded by
$\cos(\pi q(\mathbf r))$ to create a binary pulse train with local pulse width
$q(\mathbf r)$.  This can be written in the compact form
\begin{equation}
    f(\mathbf{r}) = \operatorname{H}\left[
    \cos\left(\mathbf{G\cdot r}+ 2\pi \delta(\mathbf{r})\right) - \cos(\pi q(\mathbf{r}))
    \right],
    \label{eqn:GratingModulation}
\end{equation}
where $\operatorname{H}(z)$ is the Heaviside step function defined as
\begin{equation}
    \operatorname{H}(z) \equiv 
    \begin{cases}
        0 & \quad \text{if $z <  0$}\\
        1 & \quad \text{if $z\ge 0$}
    \end{cases} .
    \label{eqn:Heaviside}
\end{equation}

\section{Experiment}\label{sec:Experiment}

\begin{figure}[htbp]
    \centering
    \includegraphics[width=0.47\textwidth]{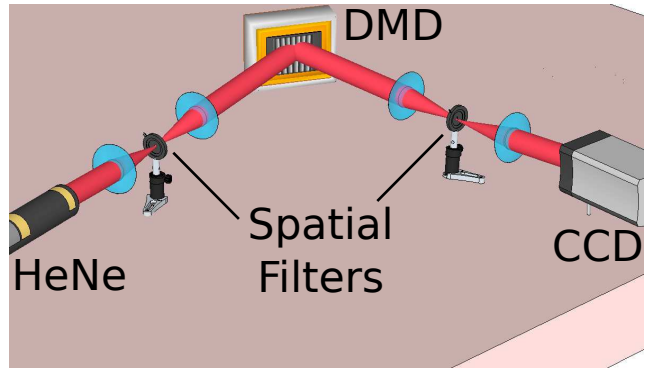}
    \caption{Experimental setup used to generate any field,
        $W(\mathbf{r_1}\mathbf{r_2})$. A fully spatially coherent plane wave is
        prepared by spatial filtering of a HeNe laser. This collimated beam is
        reflected off a CGH generated by the DMD and the desired diffracted
        order is filtered by a 4f system and imaged onto a CCD.}
    \label{fig:setup}
\end{figure}

A schematic of the experimental setup is shown in Fig.~\ref{fig:setup}. A HeNe
laser is spatially filtered using a 4f system to provide an initial coherent
plane wave incident on the DMD. The various coherent modes, $\psi_n$, are
created in rapid succession with a spatially modulated binary diffraction
grating on the DMD that gives the desired field in the first diffraction order.
A second 4f system and pinhole are used to filter out all other diffraction
orders and the resultant beam is imaged onto a CCD camera. 

The DMD is a type of micro-electronic mechanical system, commonly known as a
MEMS, that can function as an amplitude only SLM~\cite{Dudley2003}. The device
consists of a two dimensional pixelated array of micromirrors each mounted on
an individually addressed MEMS that can be in one of two positions.  In order
to use the device as a SLM, the device is aligned such that the light is
reflected and collected by the optics after the DMD if the micromirrors are in
the on position, but scattered out of the system if the mirrors are in the off
position. The device used in the experiment was a Texas Instrument DLP3000.
This device has a display resolution of $608\times 684$ pixels, a micromirror
size of $7.5\,\mathrm{\mu m}$, and the pixels can be switched at rates up to
4\,KHz which is much faster than a typical phase based
SLM~\cite{Mirhosseini2013a}.

The CCD operates at 60\,Hz, thus the detector integration time is $\tau_{det} =
1/60\,\mathrm{Hz} \approx 17\,\mathrm{ms}$. The DLP3000 DMD used in this
experiment has a switching rate of 4\,kHz, thus $\tau_s = 1/4\,\mathrm{kHz} =
250\,\mathrm{\mu s} < \tau_{det}$, which fulfills the first inequality in
Eq.~\ref{eqn:CoherenceTimes}. The bandwidth of the HeNe is 1.5\,GHz which gives
$\tau_{coh} = 1/1.5\,\mathrm{GHz} \approx 0.7\,\mathrm{ns}$ which meets the
second part of the inequality in Eq.~\ref{eqn:CoherenceTimes}.

\begin{figure}[htpb]
    \centering
    \includegraphics[width=0.47\textwidth]{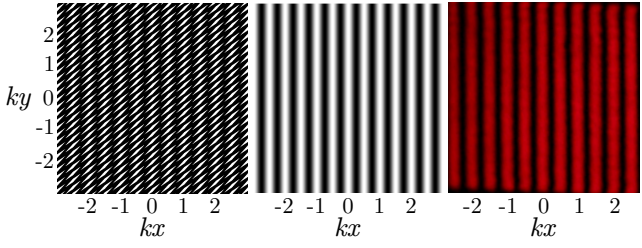}
    \caption{Interference fringes formed from the coherent superposition of two
        plane waves. Left figure shows the CGH used to generate the desired
        mode. Middle figure represents the target image while the right figure
        is an experimental image of the generated mode.}
    \label{fig:CoherentData}
\end{figure}

As a demonstration of the ability to generate a single coherent state the field
\begin{equation}
    U(\mathbf{r}) \propto e^{ik x} + e^{-ik x}
    \label{eqn:CoherentMode}
\end{equation}
was generated. This represents a coherent superposition of two plane wave
states, which form a sinusoidal interference pattern as shown in
Fig~\ref{fig:CoherentData}.

For this experiment the mode was generated using a grating with wavevector
\begin{equation}
    \mathbf G = \frac{2\pi}{\unit[25]{px}}(\mathbf{\hat{x}} + \mathbf{\hat{y}}),
\end{equation}
which represents a period of $T = \unit[25\sqrt{2}]{pixels} \approx
\unit[275]{\mu m}$ oriented at $\theta = 45^\circ$. This value of $\mathbf G$
was chosen to be large enough to allow enough separation in the Fourier plane
to allow for filtering of the 1st diffracted order with an iris.  In addition a
nonzero value was chosen for both the $x$ and $y$ components of $\mathbf G$ in
order to minimize the noise by ensuring that the diffracted order did not
overlap with any specular reflection due to the DMD's imperfect pixel
fill-fraction. The underlying grating can be seen in the left image in
Fig.~\ref{fig:CoherentData} which have the appearance of the small diagonally
oriented slivers. The plane wave transverse wavenumber was chosen to be
\begin{equation}
    k = \frac{2\pi}{\unit[100]{px}} \approx \frac{2\pi}{\unit[780]{\mu m}}.
\end{equation}
$k\ll |\mathbf G|$ and thus is slowly varying enough to allow us to use the
procedure in section~\ref{sec:BinaryGratings} to construct the CGH to create
this state. Since we are perfectly interfering 2 plane waves, the intensity
varies as $I\propto\cos^2(kx)$. Therefore $q(\mathbf r) =
\arcsin(\cos(kx))/\pi$, while $\delta(\mathbf r) = 0$.

Next we created a superposition of the plane waves $U_A = e^{ik x}$ and $U_B
= e^{-ik x}$ as before, but this time the degree of coherence between the two
beams was spatially varied, creating a partially coherent mix of modes. The
coherent modes used to represent this is given by 
\begin{equation}
    \label{eqn:PartiallyCoherentModes}
    \begin{split}
        \psi_1(\mathbf{r}) &\propto (U_A + f(\mathbf{r}) U_B)\\
        \psi_2(\mathbf{r}) &\propto (f(\mathbf{r}) U_A + U_B),
    \end{split}
\end{equation}
where the relative probability weightings are given as $p_1=p_2=1/2$, and where
$f(\mathbf{r})$ is related to the fringe visibility $V(\mathbf{r})$ by
\begin{equation}
    f(\mathbf{r}) = V(\mathbf r)/(1+\sqrt{1-V(\mathbf{r})^2}).
    \label{eqn:FringeVisibility}
\end{equation}
The intensity for this beam is
\begin{equation}
    I(\mathbf r) \propto (1-f)^2 + 4f\cos^2(kx),
\end{equation}
which is the sum of an incoherent and a coherent term which can be continuously
tuned from fully coherent ($f=1$) to incoherent ($f=0$).

\begin{figure}[htpb]
    \centering
    \includegraphics[width = 0.47\textwidth]{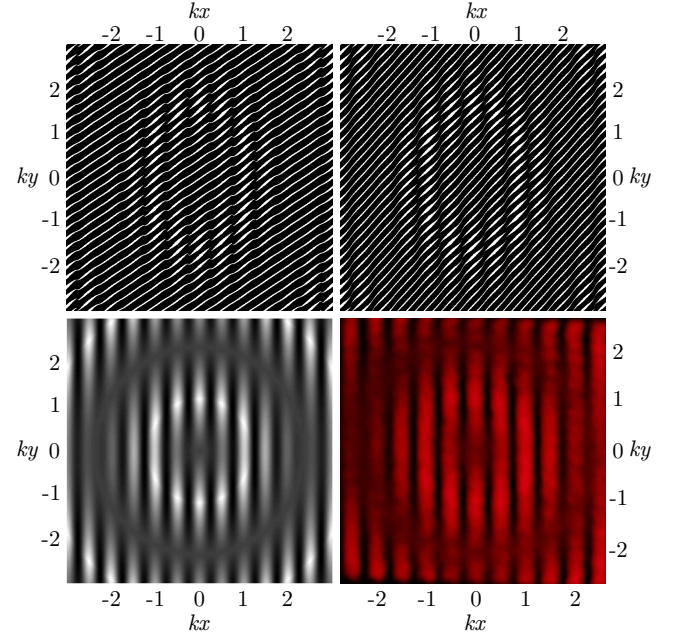}
    \caption{Interference fringes formed from superposition of two plane waves
        that are partially coherent with respect to each other. Top figures
        show the CGHs used to generate the desired modes given by
        Eq.~\ref{eqn:FringeVisibility}.  Bottom left figure represents the
        target intensity pattern, while the bottom right figure is an
        experimental image of the generated field.}
    \label{fig:PartiallyCoherent}
\end{figure}

The visibility function chosen for the experiment is given by
\begin{equation}
    V(\mathbf{r}) = |\sin(\kappa r)|,
    \label{eqn:visibility}
\end{equation}
where
\begin{equation}
    2\pi\kappa = \frac{4k}{3} = \frac{2\pi}{\unit[75]{px}}
    \approx \frac{2\pi}{\unit[580]{\mu m}}.
\end{equation}
Since $f(\mathbf r)$ was chosen to be real, Eq.~\ref{eqn:visibility} also
represents our spectral degree of coherence at $\mathbf r$. The CGHs necessary
to create the modes $\psi_1$ and $\psi_2$
(Eq.~\ref{eqn:PartiallyCoherentModes}) for this spatially varying fringe
visibility are shown in the top row of Fig.~\ref{fig:PartiallyCoherent}. The
CGH parameters are
\begin{equation}
    q(\mathbf r) = \frac{1}{\pi}\arcsin\left(\frac{\sqrt{4f\cos^2(kx)+(1-f)^2}}{I_{\text{max}}}\right),
\end{equation}
where $I_{\text{max}}$ is the maximum value of $I(\mathbf r)$ and
\begin{equation}
\begin{split}
    \delta_{1,2}(\mathbf r)
    &= \arg (\Re(\psi_{1,2})+i\Im(\psi_{1,2}))\\
    &= \arg((2\cos(kx) - (1-f)\cos(kx))\\
    &\qquad \mp i((f-1)\sin(kx))).
\end{split}
\end{equation}

\begin{figure}[htpb]
    \centering
    \includegraphics[width=0.47\textwidth]{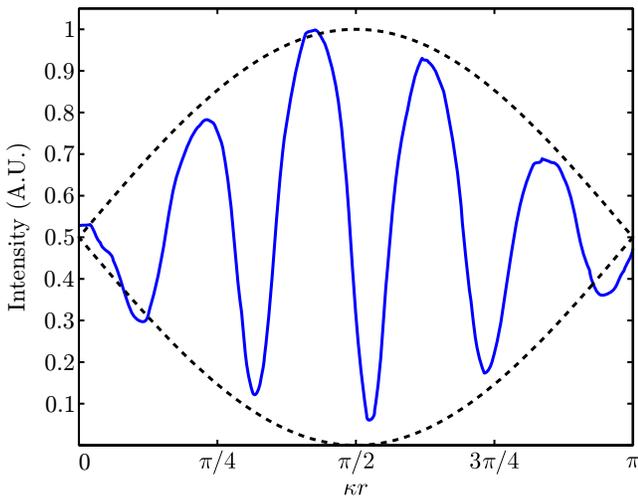}
    \caption{Plot of the intensity of the image in
        Fig.~\ref{fig:PartiallyCoherent} along the 1D slice of $r$ for
        $\theta=0$, i.e. along the $x$-axis (solid blue line). Also shown as
        the black dotted line is the theoretical envelope of the maximum and
        minimum intensities based on the intended visibility function
        $V(\mathbf r)$.}
    \label{fig:FringeVisibility}
\end{figure}

In order to compare the intended visibility given by Eq.~\ref{eqn:visibility}
with the image shown in Fig.~\ref{fig:PartiallyCoherent}, a one dimensional
slice of the intensity is plotted in Fig.~\ref{fig:FringeVisibility}. This
slice is a radial slice $r$ along the $x$ axis (i.e. at an orientation of
$\theta=0$), and is plotted over an entire period of $\sin(\kappa r)$ of the
visibility. In addition the theoretical envelope of the visibility equal to
$(1\pm V(\mathbf r))/2 = (1\pm|\sin(\kappa r)|)/2$ is plotted for comparison.
As can be seen in both the original coherent and partially coherent cases, the
intended and measured patterns are in excellent agreement with one another.

\section{Conclusions}
In this paper we have demonstrated a novel method of generating arbitrary
fields of light. Any partially coherent field that is described by the
cross-spectral density function $W(\mathbf{r_1,r_2})$ can be generated by
computing the coherent mode decomposition into an incoherent sum of coherent
modes. This incoherent mix of modes was physically realized by rapid generation
of spatial holograms on a DMD and was temporally averaged in detection.

We acknowledge Mayukh Lahiri, Joe Vornehm and Alex Radunsky for helpful
discussions. Our work was supported by the Defense Advanced Research Projects
Agency (DARPA) InPho program and OSML also acknowledges support from the
CONACyT.


\end{document}